\documentstyle[twocolumn,eqsecnum,aps]{revtex}
\begin{document}
\newcommand{\bm}[1]{\mbox{\bf #1}}
\newcommand{\bs}[2]{{#1}_{\scriptscriptstyle {\rm #2}}}
\draft
\title{Solar system tests of the equivalence principle and \\
       constraints on higher-dimensional gravity}
\author{J. M. Overduin}
\address{Department of Physics, University of Waterloo, ON, Canada
         N2L 3G1}
\address{and}
\address{Gravity Probe B, Hansen Experimental Physics Laboratory,
         Stanford University, California}
\date{\today}
\maketitle
\begin{abstract}
In most studies of equivalence principle violation by solar system
bodies, it is assumed that the ratio of gravitational to inertial mass 
for a given body deviates from unity by a parameter $\Delta$ which is 
proportional to its gravitational self-energy.  Here we inquire what
experimental constraints can be set on $\Delta$ for various solar system
objects when this assumption is relaxed.
Extending an analysis originally due to Nordtvedt,
we obtain upper limits on linearly independent combinations of $\Delta$
for two or more bodies from Kepler's third law, the position of Lagrange
libration points, and the phenomenon of orbital polarization.
Combining our results, we extract numerical upper bounds 
on $\Delta$ for the Sun, Moon, Earth and Jupiter, using observational
data on their orbits as well as those of the Trojan asteroids.
These are applied as a test case to the theory of higher-dimensional
(Kaluza-Klein) gravity.  The results are three to six orders of 
magnitude stronger than previous constraints on the theory,
confirming earlier suggestions that extra dimensions play a negligible
role in solar system dynamics and reinforcing the value of equivalence
principle tests as a probe of nonstandard gravitational theories.
\end{abstract}

\section{Introduction} \label{Sec:Int}

We investigate the consequences of possible violations of the
equivalence principle (EP) for solar system bodies whose ratio
of gravitational mass $m_g$ to inertial mass $m_i$ is given by
\begin{equation}
m_g/m_i = 1 + \Delta \; .
\label{Def_Delta}
\end{equation}
The study of this problem has a long history stretching back to Newton 
\cite{Wil93a,Dam96}.  In modern times it has been most closely associated
with Nordtvedt \cite{Nor68a,Nor68b,Nor68c,Nor70,Nor91,Nor95,Nor96},
who investigated the possibility that $\Delta$ is proportional to 
the body's {\em relative gravitational self-energy\/} $U$
\begin{eqnarray}
\Delta & = & \eta \, U \; , \label{Del_PPN} \\
U & = & -\frac{G}{c^2} \int \rho(x) \, \rho(x^{\prime})
         \frac{d^3x \, d^3x^{\prime}}{|\bm{x}-\bm{x}^{\prime}|} \, / \,
         \int \rho(x) \, d^3x \; , \nonumber
\end{eqnarray}
where $\eta$ is a universal constant made up of
parametrized post-Newtonian (PPN) parameters, 
$\eta \equiv 4\beta - \gamma - 3 - 10\xi/3 - \alpha_1 + 2\alpha_2/3 - 
2\zeta_1/3 - \zeta_2/3$ \cite{Nor68b}.  In the standard PPN gauge,
$\gamma$ is related to the amount of space curvature per unit mass,
$\beta$ characterizes the nonlinearity of the theory, $\xi$ and $\alpha_k$
refer to possible preferred-location and preferred-frame effects,
and the $\zeta_k$ allow for possible violations of momentum 
conservation (see \cite{Wil93a} for discussion).
One has $\eta=0$ in standard general relativity, where
$\gamma=\beta=1$.  In 4D scalar-tensor theories, by contrast,
$\gamma=(1+\omega)/(2+\omega)$ and
$\beta=1+\omega^{\prime}/[2(2+\omega)(3+2\omega)^2]$,
where $\omega=\omega(\phi)$ is the generalized Brans-Dicke parameter 
and $\omega^{\prime} = d\omega/d\phi$.

The relative gravitational self-energy $U$ can be calculated for
most objects in the solar system, subject to uncertainties in their
mass density profiles.  Thus, for example, $\bs{U}{S} \sim - 10^{-5}$
for the Sun, while Jupiter has $\bs{U}{J} \sim -10^{-8}$ \cite{Nor68a}.
More precise estimates are available for the Earth and Moon, 
$\bs{U}{E} = -4.64 \times 10^{-10}$ and 
$\bs{U}{M} = -1.9 \times 10^{-11}$ respectively \cite{Wil96}.
With $U$ known, constraints on EP violation (in theories of this kind)
take the form of upper limits on the PPN parameter $\eta$.
The latest such bounds (from lunar laser ranging, or LLR) are
of order $|\eta| \leq 10^{-3}$ \cite{Nor96,Wil96,Mul98,Bla99}.

There exist, however, theories of gravity in which EP violations
are not necessarily related to gravitational self-energy.
In such theories, analysis of solar system data can lead only to
upper limits on $\Delta$ for the {\em individual bodies\/} involved.
These limits will in general differ from object to object.
They will, however, typically be orders of magnitude stronger than the 
above-mentioned experimental bound on $\eta$ (since they are not diluted
by the small factor $U$).  In what follows, we obtain numerical 
constraints on $\Delta$ for the Sun, Moon, Earth and Jupiter, and
apply these as an example to higher-dimensional (Kaluza-Klein)
gravity, in which EP violations are related to the curvature
of the extra part of the spacetime manifold.

\section{Equivalence Principle Violations in Kaluza-Klein Gravity} 
\label{Sec:KK}

Theories of gravity in more than four dimensions are older
than general relativity \cite{Nor14}, and most often associated
with the names of Kaluza \cite{Kal21} and Klein \cite{Kle26} 
(for recent reviews, see \cite{Ove97,Wes99}).
The extra dimensions have traditionally been assumed to be compact,
in order to explain their nonappearance in low-energy physics.
The past few years, however, have witnessed an explosion of new interest
in {\em non\/}-compactified theories of higher-dimensional gravity
\cite{Wes96,Ark98,Ran99}.  In such theories the dimensionality
of spacetime may in principle manifest itself at experimentally accessible
energies.  We examine here the theory of Kaluza-Klein gravity \cite{Ove97},
and focus on the prototypical five-dimensional (5D) case, although the
extension to higher dimensions is straightforward in principle.

In 5D Kaluza-Klein gravity, objects such as stars and planets are modelled
by a static, spherically-symmetric analog of the 4D Schwarzschild solution
known as the {\em soliton metric\/}.  This may be written
(following \cite{Gro83}, but switching to nonisotropic form, 
and defining $a\equiv 1/\alpha$, $b\equiv \beta/\alpha$ and 
$M_{\ast}\equiv 2m$)
\begin{eqnarray}
dS^{\,2} & = & A^a dt^2 - A^{-a-b} dr^2 - A^{1-a-b} \, \times \nonumber \\
& & r^2 \! \left( d\theta ^2 + \sin ^2\theta d\phi ^2 \right) - A^b dy^2 \; ,
\label{Sol_Metric}
\end{eqnarray}
where $y$ is the fifth coordinate,
$A(r) \equiv 1-2M_{\ast}/r$, $M_{\ast}$ is a parameter related to the mass 
of the object at the center of the geometry, and the constants $a,b$ 
satisfy a consistency relation $a^2+ab+b^2 = 1$, or
\begin{equation}
a = -b/2 \pm (1 - 3b^2/4)^{1/2} \; ,
\label{a_of_b}
\end{equation}
which follows from Einstein's field equations.
Equation~(\ref{Sol_Metric}) reduces to the 4D Schwarzschild solution 
on hypersurfaces $y=$~const as $b \rightarrow 0$ and $a \rightarrow +1$.

Solitons differ in interesting ways from 4D black holes, as discussed by many
authors \cite{Ove97,Wes99,Gro83,Soliton,MoreSoliton}.
Of most relevance for our purposes is the fact that a soliton's gravitational
mass (as identified from the asymptotic behaviour of $\bs{g}{00}$) is given
by $m_g = a M_{\ast}$, while its inertial mass (as obtained using the
Landau-Lifshitz energy-momentum pseudotensor) turns out to be
$m_i = (a+b/2) M_{\ast}$.  Using equations~(\ref{Def_Delta}) and
(\ref{a_of_b}), it follows that
\begin{equation}
\Delta = \mp b/2 \, ( 1 - 3b^2/4 )^{-1/2} .
\label{Del_KK_roots}
\end{equation}
It is important to recognize that $b$ is not a universal constant
like $\eta$, but depends in principle on local physics \cite{MoreSoliton}
and may vary from soliton to soliton.  (Birkhoff's theorem in its usual 
form does not hold in higher-dimensional general relativity \cite{Birk}.) 
To constrain the theory, one therefore hopes to apply as many tests as 
possible to a given astrophysical system.  

The classical tests of general relativity have been worked out
for the soliton metric.  Data from long-baseline radio interferometry, 
ranging to the Mars Viking lander, and the perihelion precession of Mercury
imply that $|b| \leq 0.07$ in the solar system \cite{Kal95,Liu00}.
In what follows, therefore, we keep only terms of second order
or lower in $b$.  We also drop the negative roots in equation~(\ref{a_of_b})
[ie, the positive roots in equation~(\ref{Del_KK_roots})],
in order to eliminate the possibility of negative gravitational and/or
inertial mass, and to ensure that the 4D Schwarzschild solution
is recovered (for $y=$~const) as $b \rightarrow 0$ \cite{Ove97}.
This leaves us with 
\begin{equation}
\Delta = -b/2 \; ,
\label{Del_KK}
\end{equation}
which may be compared with equation~(\ref{Del_PPN}) for PPN-type theories.
In Kaluza-Klein gravity, experimental constraints on EP violation 
translate directly into upper bounds on the metric parameter $b$, 
which characterizes the departure from flatness of the fifth dimension
in the vicinity of the central mass.

\section{The Modified Third Law of Kepler} \label{Sec:Kep}

Let us now proceed to see what experimental constraints can be placed on
the parameter $\Delta$ for solar system bodies.  Given any inverse-square
law central force ($F=-\kappa/r^2$), it may be shown that the two-body 
problem admits as solutions elliptical orbits satisfying
\begin{equation}
\tau = 2\pi a^{3/2} \sqrt{\mu/\kappa} \; ,
\label{Kep3}
\end{equation}
where $\tau$ is the orbital period, $a$ the semi-major axis, and
$\mu \equiv m_{i,1}m_{i,2}/(m_{i,1}+m_{i,2})$ the reduced mass of
the system \cite{Gol80}.  (Here $m_{i,k}$ is the inertial
mass of the $k$th body.)  We are concerned in this paper with the case in 
which $\kappa=G m_{g,1} m_{g,2}$, where the gravitational masses $m_{g,k}$ 
are related to the inertial ones $m_{i,k}$ by equation~(\ref{Def_Delta}).
Rearranging equation~(\ref{Kep3}), one then obtains the following 
modified form of Kepler's third law
\begin{equation}
G(m_1+m_2+m_2\Delta_1+m_1\Delta_2) = \omega^2 a^3 \; ,
\label{mod_Kep3}
\end{equation}
where $\omega \equiv 2\pi/\tau$ is the mean orbital angular frequency,
and $m_k$ refers (here and elsewhere in the remainder of this paper) to
the {\em gravitational\/} mass of the $k$th body.  Equation~(\ref{mod_Kep3})
was first obtained by Nordtvedt using a somewhat different argument \cite{Nor68a}.

If all bodies have the {\em same\/} value of $\Delta$, we will not see 
violations of the EP.  The common part of $\Delta_1$ and $\Delta_2$ can
be absorbed into a rescaled gravitational constant.  One way to see this
explicitly is to rewrite equation~(\ref{mod_Kep3}) in the form
\begin{equation}
G(1+\Delta_1)(m_1+m_2) + Gm_1(\Delta_2-\Delta_1) = \omega^2 a^3 \; .
\label{mod_Kep4}
\end{equation}
There are two modifications to Kepler's third law here:
a rescaling of the value of $G$ in the first term, and a completely
new second term, which depends only on the difference $\Delta_2-\Delta_1$.
This latter term is a clear manifestation of EP violation in the system.

Nevertheless, we can put experimental constraints on the values of 
$\Delta_1$ and $\Delta_2$, even in cases where they are equal.  To see this, 
divide equation~(\ref{mod_Kep4}) through by $G m_1$ and rearrange to obtain
\begin{eqnarray}
\frac{\omega^2 a^3}{G m_1} - \left( 1 + \frac{m_2}{m_1} \right) 
   & = & (\Delta_2 - \Delta_1) \nonumber \\
& & + \Delta_1 \left( 1 + \frac{m_2}{m_1} \right) \; .
\label{mod_Kep5}
\end{eqnarray}
To within experimental uncertainty, of course, the left-hand side of 
this equation vanishes for planetary two-body systems; this is a statement 
of Kepler's third law.  (More accurately, we may say that in standard theory, 
where $\Delta_1$ and $\Delta_2$ are assumed to vanish, a statistical 
best-fit value is chosen for $G m_1$ in such a way as to {\em force\/} the
left-hand side as close to zero as possible for all systems observed.)
Even when $(\Delta_2-\Delta_1)=0$, therefore, observation imposes an upper
bound on the possible size of the last term on the right-hand side.
This reflects the fact that we have excellent experimental data, 
not only on possible violations of the EP, but also on the value of 
$G m_1$ (as well as $\omega,a$ and $m_2/m_1$) for most solar system objects. 
(The situation is less satisfactory for $G$ and $m_1$ considered 
individually \cite{Mon94}, but these quantities are not needed here.)

We can use this extra knowledge to set limits on the {\em sum\/},
rather than the difference of $\Delta_1$ and $\Delta_2$.
In combination with more purely EP-based constraints from the three-body 
problem, this will allow us to extract limits on the individual terms 
$\Delta_1$ and $\Delta_2$.  We therefore collect terms in $\Delta_1$ 
and $\Delta_2$ and recast equation~(\ref{mod_Kep5}) in a form more
suitable for contact with observation
\begin{eqnarray}
\frac{m_2}{m_1} \, \Delta_1 + \Delta_2 & = & \left( \frac{m_s}{m_1} \right)
   \left( \frac{\omega}{k} \right)^2 \left( \frac{a}{A} \right)^3
   \nonumber \\
& & - \left( 1 + \frac{m_2}{m_1} \right) \; ,
\label{Kep_con}
\end{eqnarray}
where $k$ is the Gaussian constant, $a$ the semi-major axis of the orbit,
and $A$ the length of the astronomical unit (AU).  The quantities $k$ and
$A$ are related via $k^2 A^3\equiv G\bs{m}{S}$ to the gravitational mass
of the Sun (which is actually inferred in practice from a statistical fit
to experimental data on $A$).  The length of the AU ($A=1.5 \times 10^{11}$~m)
is currently known from planetary ranging data to better than $\pm 20$~m 
\cite{Wil89}.

To within experimental accuracy, the right-hand side of 
equation~(\ref{Kep_con}) vanishes (as remarked above; Kepler's third law).
Our upper limits on the left-hand side are then just the sum of the 
relevant uncertainties in $m_s/m_1$, $\omega$, $a$, $A$ and $m_2/m_1$
($k$ is an exact quantity).

For the Sun-Jupiter system ($m_1=\bs{m}{S}$ and $m_2=\bs{m}{J}$),
$\omega (=0.53$~rad/yr) has an uncertainty which may be conservatively
estimated at 1~arcsec/250~yr \cite{Sta92}.  The mean Sun-Jupiter 
distance $(=5.2$~AU) is likewise uncertain to about 200~km, based on 
post-fit residuals to all observational data \cite{Wil89} (not individual
ranging measurements to Jupiter, which are good to as little as 
3~km \cite{Fol95}).  And the mass ratio $\bs{m}{S}/\bs{m}{J} (=1047$)
is known to $\pm 0.0005$ \cite{Wil89}.  Combining these numbers, 
we find that
\begin{equation}
| \frac{\bs{\Delta}{S}}{1047} + \bs{\Delta}{J} | \leq 8 \times 10^{-7} \; .
\label{Kep_Con1}
\end{equation}
The bulk of this is due to the uncertainty in $a$, with a small
remainder coming from uncertainty in $\omega$.  The uncertainties in $A$
(that is, in $G\bs{m}{S}$) and $\bs{m}{J}/\bs{m}{S}$ are unimportant 
by comparison.

For the Earth-Moon system ($m_1=\bs{m}{E}, m_2=\bs{m}{M}$),
$\omega (=0.23$~rad/day) has an uncertainty of less than 0.03~arcsec/century
\cite{Wil93b}.  The mass ratio $\bs{m}{S}/(\bs{m}{E}+\bs{m}{M}) 
(=3.3 \times 10^5$) is known to better than $\pm 0.002$ \cite{Dic94}, while 
$\bs{m}{E}/\bs{m}{M} (=81$) is good to $\pm 0.00002$ \cite{Kon98}. 
The mean Earth-Moon distance $(=3.8 \times 10^5$~km) has been 
established to within 0.8~m by LLR data \cite{Dic94}.  Let us take
the uncertainty in the length of the semi-major axis $a$ to be less
than twice this figure, or 1.6~m.  We then obtain
\begin{equation}
| \frac{\bs{\Delta}{E}}{81} + \bs{\Delta}{M} | \leq 1 \times 10^{-8} \; .
\label{Kep_Con2}
\end{equation}
In this case, while uncertainty in $a$ is still responsible for about
three-quarters of the total, there are also substantial contributions from
uncertainties in $\bs{m}{S}/\bs{m}{E}$ and $\bs{m}{M}/\bs{m}{E}$.
The decrease in importance of error in $a$ (relative to the Sun-Jupiter case)
may be attributed to the high quality of the LLR data.

\section{Migration of the Stable Lagrange Points and the Trojan
         Asteroids} \label{Sec:Lag}

We turn next to the effects of EP violation on three-body motion in
the solar system, beginning with the problem in which a test body
is located at one of the stable Lagrange points ($L=L_4$ or $L_5$) in the 
orbit of one massive body $M_2$ (a planet or moon, say) around another 
($M_1$).  Nordtvedt \cite{Nor68a} has shown that the location of $L$ is
sensitive to $\Delta_1 \neq 0$ for the largest of the three masses.
We follow his notation, which is based on the assumption of circular
orbits, but extend the calculation to cases in which all {\em three\/} 
objects can in principle display EP violations of the form~(\ref{Def_Delta}).

The point $L$ is located at points where radial acceleration toward the
$M_1$-$M_2$ center of mass ($C$, say) is precisely balanced by centrifugal
acceleration, and where there is no net acceleration perpendicular to this
direction.  The situation is illustrated in Fig.~1, where $r_1$ and 
$r_2$ denote the distances from $L$ to $M_1$ and $M_2$ respectively,
while $r$, $R_1$ and $R_2$ are the distances of $L$, $M_1$ and $M_2$ 
from $C$.  It is straightforward to show \cite{Nor68a} that
\begin{equation}
(r_1/r_2)^3 = (1+\Delta_1)/(1+\Delta_2) \; .
\end{equation}
The effect of the $\Delta$-terms, in other words, is to ``de-equilateralize''
the triangle made up by $L$, $M_1$ and $M_2$.

It is further possible to solve explicitly for $r_1$ and $r_2$; one finds
that
\begin{eqnarray}
\left[ r_1/(R_1+R_2)\right] ^3 & = & (1+\bs{\Delta}{L})/(1+\Delta_2) \nonumber \\
\left[ r_2/(R_1+R_2)\right] ^3 & = & (1+\bs{\Delta}{L})/(1+\Delta_1) \; .
\end{eqnarray}
Denoting the unperturbed values of $r_1$ and $r_2$ by $R$ ($\equiv R_1+R_2$),
and assuming that the $\Delta$-terms are much less than unity, we see that 
they lead to small movements of $L$ toward (or away from) $M_1$ and $M_2$, 
as follows
\begin{eqnarray}
\delta r_1 & \approx & - (\Delta_2-\bs{\Delta}{L}) (R_1+R_2) /3 \nonumber \\
\delta r_2 & \approx & - (\Delta_1-\bs{\Delta}{L}) (R_1+R_2) /3 \; .
\label{del_rs}
\end{eqnarray}
In the limits $\bs{\Delta}{L} \rightarrow 0$ and $\Delta_2 \rightarrow 0$, 
these results reduce to those given in \cite{Nor68a}.

Experimentally, one could use these results to obtain constraints on
$(\Delta_k-\bs{\Delta}{L})$ by ranging to a satellite at $L$ and looking for 
nonzero values of $\delta r_k$ ($k=1$ or $2$).  In the case of the 
Earth-Moon system ($M_1=$~Earth, $M_2=$~Moon) this would require an 
artificial satellite such as the SOHO spacecraft (which is however 
located at $L_1$ rather than $L_4$ or $L_5$).  In the case of the 
Sun-Jupiter system ($M_1=$~Sun, $M_2=$~Jupiter) one could instead make 
use of the larger Trojan asteroids, of which 413 have now been identified
\cite{Bur99}, and some $2 \times 10^6$ estimated to exist with radii
over 1~km \cite{Lev97}.  Such a procedure would however be hindered 
by low signal-to-noise levels and uncertainties involving factors 
such as surface topography.

An observational quantity offering greater promise is the angular position
$\theta$ of the stable Lagrange point relative to the line through $M_1$ and 
$M_2$ (see Fig.~1).  For different values of $r_1$ and $r_2$, one will 
measure different angles $\delta\theta$.  From the law of sines
\begin{equation}
\tan ( \theta + \delta\theta ) = \frac{ (R-\delta r_1) \sin\beta }
                                 {R_1 - (R-\delta r_1) \cos\beta } \; ,
\end{equation}
where $\beta$ may be determined from the law of cosines
\begin{eqnarray}
(R-\delta r_2)^2 & = & R^2 + (R-\delta r_1)^2 \nonumber \\
   & & - 2R (R-\delta r_1) \cos\beta \; .
\end{eqnarray}
Differentiating, and linearizing in the small parameters $\delta r_1$ 
and $\delta r_2$, one can show that
\begin{equation}
\delta\theta = \frac{ (2R_1+R_2) \, \delta r_1 - (R_1+2R_2) \, \delta r_2 }
                    { \sqrt{3} \, (R_1^2 + R_1R_2 + R_2^2) } \; .
\end{equation}
Inserting equations~(\ref{del_rs}) into this expression, we obtain the
general result
\begin{eqnarray}
\delta\theta & = & \frac{ (R_1+R_2) }{ 3\sqrt{3} \, (R_1^2+R_1R_2+R_2^2) } 
   \left[ (R_1+2R_2)(\Delta_1-\bs{\Delta}{L}) \right. \nonumber \\
& & \hspace{25mm} \left. - \, (2R_1+R_2)(\Delta_2-\bs{\Delta}{L}) \right] \; .
\end{eqnarray}
That this is a pure EP-violating effect can be seen from the fact that
only {\em differences\/} of $\Delta$-terms appear.  In cases where
$\Delta_1=\Delta_2=\bs{\Delta}{L}$, there is no migration of the Lagrange 
point and $\delta\theta =0$.

Let us now specialize to cases in which $M_1$ is much more massive 
than $M_2$, so that $R_2 \gg R_1$.  (This is not a serious limitation
as it describes all situations of interest in the solar system.)
We also follow Nordtvedt \cite{Nor68a} in assuming that the test object
at $L$ differs from the much larger masses $M_1$ and $M_2$ in such a way 
that $\bs{\Delta}{L} \ll \Delta_1$ and $\bs{\Delta}{L} \ll \Delta_2$.
This leaves us with
\begin{equation}
\delta\theta = \frac{1}{3\sqrt{3}} ( 2\Delta_1 - \Delta_2 ) \; ,
\label{Mig_Ang}
\end{equation}
a result that can be used to set constraints on $(2\Delta_1-\Delta_2)$
if we have experimental data on any possible angular shift of the 
Lagrange points from their expected locations 60 degrees behind and 
ahead of Jupiter.

It happens that we do have good observations of a number of Trojan
asteroids, the first of which was discovered over ninety years ago.
To locate their mean angular position with sufficient precision for EP
tests is however a significant challenge because these objects undergo 
{\em librations\/} about the Lagrange points, with periods that are 
typically a good deal longer --- of order 150~yr \cite{Ore88} --- than
the timescale over which they have been observed.  Only twelve Trojans 
(588~Achilles, 911 Agamenon, 1404~Ajax, 1173~Anchises, 1172~Aneas, 
1437~Diomedes, 624~Hector, 659~Nestor, 1143~Odysseus, 617~Patroclus, 
884~Priamus and 1208~Troilus) have been observed for sixty years or more.  
So one has to fit orbits to an incomplete arc of observations.  Moreover
the older observational data will be subject to larger random scatter than
more recent measurements.  There are also several potential sources of 
systematic error, including observational selection effects and the nonuniform 
distribution of asteroids, which may not necessarily cancel themselves 
out as the libration centers follow Jupiter around the sky.

The problem has nevertheless been seriously tackled in a pair of papers
by Orellana and Vucetich \cite{Ore88,Ore93}.  (The authors set out to obtain 
constraints on the PPN parameter $\eta$, but we may relate this simply
to $\delta\theta$ by a constant factor, given as $\delta\theta = 1.26 
\eta$~arcsec in \cite{Ore93}.)  A formal least-squares statistical fit
(including a model for systematic errors) produces the rather surprising
result $\delta\theta = -0.21 \pm 0.09$~arcsec, which could perhaps be
taken as a potential EP-violating signal.  A more conservative procedure,
however, leads to a final value of
$\delta\theta = -0.18 \pm 0.15$~arcsec, 
which is consistent with no signal.  Based on this latter limit we adopt
the upper bound $|\delta\theta| \leq 0.33$~arcsec.  This compares reasonably
with planetary angular measurements, which are typically accurate to between
0.1 and 1~arcsec \cite{Wil89}.  In combination with equation~(\ref{Mig_Ang})
this bound leads to the constraint
\begin{equation}
| 2\bs{\Delta}{S} - \bs{\Delta}{J} | \leq 8 \times 10^{-6} \; .
\label{Lag_Con}
\end{equation}
This is an order of magnitude weaker than the limit~(\ref{Kep_Con1})
on the sum $\bs{\Delta}{S}/{1047} + \bs{\Delta}{J}$ for the
Sun-Jupiter system from Kepler's third law, and almost {\em three\/} 
orders of magnitude weaker than the corresponding limit~(\ref{Kep_Con2})
for the Earth-Moon system.  Nevertheless equation~(\ref{Lag_Con}) will
prove useful in constraining the value of $\bs{\Delta}{J}$.

\section{Orbital Polarization and the Lunar Nordtvedt Effect}
\label{Sec:Lunar}

We turn next to the best-known probe of EP violations in the solar system,
the polarization of a two-body orbit in the field of a third body, also
known variously as the Nordtvedt Effect and the Lunar E\"otv\"os experiment.
This was investigated by Nordtvedt in the context
of the Earth-Moon-Sun system \cite{Nor68a,Nor68b,Nor68c} and the 
Sun-Earth-Jupiter system \cite{Nor70}, and has been the subject of numerous
studies since \cite{Wil93a,Dam96,Nor91,Nor95,Nor96,Wil96,Mul98}.
We follow the approach of Will \cite{Wil93a}, but extend his notation to
general three-body systems.  The situation is illustrated in Fig.~2.

Including their mutual attraction (but neglecting any quadrupole or
higher moments), the two massive bodies $M_1$ and $M_2$ described by
equation~(\ref{Def_Delta}) will fall toward $M_3$ with different 
accelerations given by
\begin{eqnarray}
\bm{a}_1 & = & -(1+\Delta_1) (m_3\bm{x}_1/r_1^3 - m_2\bm{x}/r^3 ) \; , 
   \nonumber \\
\bm{a}_2 & = & -(1+\Delta_2) (m_3\bm{x}_2/r_2^3 + m_1\bm{x}/r^3 ) \; , 
\end{eqnarray}
where $\bm{x}_1$ and $\bm{x}_2$ are the 3-vectors connecting $M_3$ to
$M_1$ and $M_2$ respectively, and $\bm{x}$ is the 3-vector from $M_1$ to
$M_2$.  (The lengths of these vectors are $r_1, r_2$ and $r$ respectively.)
The {\em relative\/} acceleration $\bm{a}\equiv\bm{a}_2-\bm{a}_1$
between $M_1$ and $M_2$ may then be written
\begin{eqnarray}
\bm{a} & = & -m_{\ast} \frac{\bm{x}}{r^3} + (\Delta_1 - \Delta_2) \, m_3
         \frac{\bm{x}_1}{r_1^3} \nonumber \\ 
       & & + (1+ \Delta_2) \, m_3 \left( \frac{\bm{x}_1}{r_1^3} - 
         \frac{\bm{x}_2}{r_2^3} \right) \; ,
\end{eqnarray}
where $m_{\ast} \equiv m_1(1+\Delta_2)+m_2(1+\Delta_1)$.
The first term is the standard Newtonian acceleration, which gives rise to the
circular (unperturbed) orbit of $M_2$ around $M_1$, assumed here to be coplanar
with $M_3$.  The second, ``Nordtvedt term'' will be treated as a perturbation 
of this orbit.  (We follow Will \cite{Wil93a} in neglecting the third, tidal
term at this stage.)  The acceleration $\bm{a}=d^2\bm{x}/dt^2$ and angular 
momentum $\bm{h}=\bm{x}\times(d\bm{x}/dt)$ satisfy
\begin{eqnarray}
d^2r/dt^2 & = & \bm{x $\cdot$ a}/r + h^2/r^3 \; , \nonumber \\
d\bm{h}/dt & = & \bm{x} \times \bm{a} \; .
\end{eqnarray}
Defining $\delta a \equiv (\Delta_1-\Delta_2) m_3 / r_1^2$,
linearizing about the unperturbed orbit (so that $r \equiv r_0 + \delta r$
and $h \equiv h_0 + \delta h$) and solving the resulting pair of differential
equations, we find the following periodic oscillation in distance 
between $M_1$ and $M_2$
\begin{equation}
\delta r = (\Delta_1 - \Delta_2) \, \bs{A}{EP}
   \cos D \; ,
\label{Nor_effect}
\end{equation}
where $\bs{A}{EP}$, the EP-violating amplitude, 
is given by
\begin{equation}
\bs{A}{EP} \equiv \left[ 
   \frac{1+2\omega_2/(\omega_2-\omega_1)}{2 \, (\omega_2/\omega_1)-1} 
   \right] r_1 \; ,
\label{AEP}
\end{equation}
and $D \equiv (\omega_2-\omega_1)\,t$ is the synodic phase.
Here $\omega_1$ is the orbital angular frequency of $M_1$ about $M_3$, 
$\omega_2$ is the orbital angular frequency of $M_2$ about $M_1$,
and we have made use of the relations $m_{\ast}/r_0^2 = \omega_2^2 r_0$,
$m_3/r_1^2 = \omega_1^2 r_1$ and $h_0/r_0^2 = \omega_2$.

Equation~(\ref{Nor_effect}) represents a polarization, or alignment of 
the orbit of $M_2$ about $M_1$ along the direction either toward 
(if $\Delta_1 > \Delta_2$) or away from $M_3$ (if $\Delta_1 < \Delta_2$).
For the Earth-Moon system ($M_1=$~Earth, $M_2=$~Moon, $M_3=$~Sun), we use 
$\omega_1=1.991 \times 10^{-7}$~rad/s, $\omega_2=2.662 \times 10^{-6}$~rad/s
and $r_1=1.496 \times 10^{11}$~m to obtain $\bs{A}{EP}=1.84 \times 10^{10}$~m.
Due to amplification by tidal effects, however, this figure is 
thought to be a considerable underestimate \cite{Wil93a,Nor95}.
More refined calculations, both analytical and numerical, now
indicate a final value of $\bs{A}{EP}=2.9 \times 10^{10}$~m 
\cite{Dam96,Wil96,Mul98}.

All these amplitudes are, of course, far larger than any observed
anomalous fluctuations in the Earth-Moon distance at this frequency.
(The largest known term at the synodic frequency has an amplitude of
approximately 100~km, but can be modelled so accurately that it does 
not contribute to the uncertainty in $\delta r$.)  Indeed, analysis of LLR
data now constrains any such fluctuations to be less than 1.3~cm in size
\cite{Nor96,Wil96,Mul98}.  Using the latest above-mentioned value 
for $\bs{A}{EP}$, we infer an upper bound
\begin{equation}
|\bs{\Delta}{E} - \bs{\Delta}{M}| \leq 4.4 \times 10^{-13} \; .
\label{Nor_Con1}
\end{equation}
For PPN-type theories, in which $\bs{\Delta}{E} = \eta \bs{U}{E}$ and
$\bs{\Delta}{M} = \eta \bs{U}{M}$ with the values of $\bs{U}{E}$
and $\bs{U}{M}$ given in \S~\ref{Sec:Int}, this leads to the
constraint $|\eta | \leq 1.0 \times 10^{-3}$ \cite{Nor96,Wil96},
which is currently the strongest limit on this parameter.
(It has been weakened only slightly, to $|\eta | \leq 1.3 \times 10^{-3}$,
by a recent study of possible masking by composition-dependent effects
\cite{Bla99}.)

\section{Orbital Polarization and the Solar Nordtvedt Effect}
\label{Sec:Solar}

One can also apply equation~(\ref{Nor_effect}) to the Sun-Earth
system with Jupiter as the third mass ($M_1=$~Sun, $M_2=$~Earth, 
$M_3=$~Jupiter), as suggested in \cite{Nor70}.  This might be dubbed a
{\em solar\/} E\"otv\"os experiment since it is the difference in 
accelerations of the Earth and Sun, and the size of anomalous variations in
the solar-terrestrial distance, that are of interest.  For this case we have
$\omega_1=1.679 \times 10^{-8}$~rad/s, $\omega_2=1.991 \times 10^{-7}$~rad/s
and $r_1=7.783 \times 10^{11}$~m, resulting in an uncorrected Nordtvedt 
factor of $\bs{A}{EP}=1.09 \times 10^{11}$~m.  

A realistic upper limit on anomalous fluctuations in the Sun-Earth 
distance at the relevant frequency (corresponding to a period of 1.09~yr)
might be several times the uncertainty in the length of the AU ($\pm 20$~m);
we take here a value of 100~m.  (There are ordinary classical perturbations
of amplitude $\sim 2000$~km at this frequency \cite{Nor70}, but these can 
be compensated for very accurately, since the ratio of Jovian to solar 
mass is known to better than a part in $10^6$ \cite{Wil89}.)  Using 
equations~(\ref{Nor_effect}) and (\ref{AEP}), we then obtain
\begin{equation}
|\bs{\Delta}{S} - \bs{\Delta}{E}| \leq 9 \times 10^{-10} \; ,
\label{Nor_Con2}
\end{equation}
which is some {\em four orders of magnitude\/} stronger than the 
constraint on $\bs{\Delta}{S}$ and $\bs{\Delta}{J}$ obtained from the
Trojan asteroids in equation~(\ref{Lag_Con}).  This is a reflection of two
things: the uncertainties discussed in \S~\ref{Sec:Lag} in connection with
the motion of these bodies, and the fact that our measurements of the distance
to the Sun, like that to the Moon, rest ultimately on the great precision
of ranging data (in this case, from the Viking lander rather than the 
Apollo retroflectors).

To complete the set of constraints, we must finally apply the modified
Kepler's third law, equation~(\ref{Kep_con}), to our last pair of bodies.
For the Sun-Earth system ($m_1=\bs{m}{S}$ and $m_2=\bs{m}{E}$),
uncertainty in $\omega (=2\pi$~rad/yr) will be insignificant.  For 
definiteness one could perhaps use the total drift error in the Earth's 
inertial mean longitude, which is 0.003~arcsec/century \cite{Wil89}.
As in the Earth-Moon case, let us take the uncertainty in semi-major axis 
$a$ (here 1~AU) to be 40~m, or twice that in the length of the AU itself.
Uncertainty in the mass ratio $\bs{m}{E}/\bs{m}{S} (=1/334,000)$
may be obtained from the limits
$\delta(\bs{m}{S}/(\bs{m}{E}+\bs{m}{M})) \leq 0.002$ \cite{Dic94} and
$\delta(\bs{m}{E}/\bs{m}{M}) \leq 0.00002$ \cite{Kon98}, giving a figure
of $\delta(\bs{m}{E}/\bs{m}{S})=2 \times 10^{-14}$.  Combining these 
numbers, we find that
\begin{equation}
| \frac{\bs{\Delta}{S}}{334,000} + \bs{\Delta}{E} | \leq 7 \times 10^{-9} \; .
\label{Kep_Con3}
\end{equation}
This is due almost entirely to the uncertainty in $\bs{m}{E}/\bs{m}{S}$ ---
small though that is --- with the tiny remainder coming from uncertainty 
in the values of $a$ and $A$.  

\section{Independent Limits on $\Delta$} \label{Sec:Ind}

We now proceed to extract individual constraints on the parameters
$\Delta_k$ by combining equations~(\ref{Kep_Con1}) and (\ref{Lag_Con}),
(\ref{Kep_Con2}) and (\ref{Nor_Con1}), and (\ref{Nor_Con2}) and
(\ref{Kep_Con3}) respectively.  All three pairs of inequalities 
may be written in the form
\begin{eqnarray}
|\Delta_1 + c_1 \Delta_2| & \leq & \epsilon_1 \; , \nonumber \\
|\Delta_1 - c_2 \Delta_2| & \leq & \epsilon_2 \; ,
\label{Pair}
\end{eqnarray}
where $c_1,c_2,\epsilon_1$ and $\epsilon_2$ are positive.
We wish to obtain upper limits on the quantities $\Delta_k$.

Squaring equations~(\ref{Pair}), taking their sum and difference,
and solving the resulting pair of quadratic equations, we find that
\begin{eqnarray}
|\Delta_1| & \leq & \frac{|(c_1+c_2)(\epsilon_1 \mp \epsilon_2) -
                           (c_1-c_2)(\epsilon_1 \pm \epsilon_2)|}
                         {2(c_1+c_2)} \; , \nonumber \\
|\Delta_2| & \leq & \frac{|\epsilon_1 \pm \epsilon_2|}{c_1+c_2} \; ,
\label{Ind_Cons}
\end{eqnarray}
where the signs must be evaluated together (ie, if the upper sign is 
selected for $\Delta_1$, then it must be selected for $\Delta_2$ as well).

We apply these results to the Sun-Jupiter system (1={\sc s},2={\sc j}) by 
substituting $c_1=1047$ and $\epsilon_1=8 \times 10^{-4}$ from 
equation~(\ref{Kep_Con1}) (Kepler's modified third law), and
$c_2=0.5$ and $\epsilon_2=4 \times 10^{-6}$ from equation~(\ref{Lag_Con})
(migration of the stable Lagrange point) to obtain
\begin{eqnarray}
|\bs{\Delta}{S}| & \leq & 5 \times 10^{-6} \; ,
   \label{Del_Sun1} \\
|\bs{\Delta}{J}| & \leq & 8 \times 10^{-7} \; .
   \label{Del_Jup}
\end{eqnarray}
Similarly, for the Earth-Moon system (1={\sc e},2={\sc m}),
equation~(\ref{Kep_Con2}) gives $c_1=81$ and $\epsilon_1=8 \times 10^{-7}$
(Kepler's modified third law), while
equation~(\ref{Nor_Con1}) gives $c_2=1$ and $\epsilon_2=4 \times 10^{-13}$
(lunar Nordtvedt effect), so that
\begin{eqnarray}
|\bs{\Delta}{E}| & \leq & 1 \times 10^{-8} \; ,
   \label{Del_Ear1} \\
|\bs{\Delta}{M}| & \leq & 1 \times 10^{-8} \; .
   \label{Del_Moo}
\end{eqnarray}
Finally, for the Sun-Earth system (1={\sc s},2={\sc e}),
we take $c_1=334,000$ and $\epsilon_1=2 \times 10^{-3}$ from
equation~(\ref{Kep_Con3}) (Kepler's law), while $c_2=1$ and 
$\epsilon_2=9 \times 10^{-10}$ from equation~(\ref{Nor_Con2})
(solar Nordtvedt effect), so that
\begin{eqnarray}
|\bs{\Delta}{S}| & \leq & 8 \times 10^{-9} \; ,
   \label{Del_Sun2} \\
|\bs{\Delta}{E}| & \leq & 7 \times 10^{-9} \; .
   \label{Del_Ear2}
\end{eqnarray}
The limit~(\ref{Del_Sun2}) obtained for the Sun from the solar Nordtvedt
effect is nearly three orders of magnitude stronger than that derived from
the Trojan asteroids, equation~(\ref{Del_Sun1}).  This is perhaps not 
surprising, given the discussion in \S~\ref{Sec:Lag} above.  But the fact
that the lunar and solar Nordtvedt effects lead to {\em equally\/} strong 
constraints on $\bs{\Delta}{E}$ for the Earth --- equations~(\ref{Del_Ear1})
and (\ref{Del_Ear2}) respectively --- is somewhat unexpected, in view of the
widespread belief that LLR data provides by far the strongest probe of EP 
violations in the solar system.  This may be partly explained by the fact
that the strength of both Nordtvedt effect-based limits ultimately derives
from ranging measurements, as discussed in \S~\ref{Sec:Solar}.

Statistical effects may also play a role here.  As uncertainties in
our orbital and other parameters ($\chi$, say), we have in each case used 
residuals from published fits to a fixed number of solution parameters.
These fits, however, {\em do not generally incorporate\/} all the $\chi$
(they are typically sensitive to at most a single EP-violating parameter, 
$\eta$).  We have, in other words, relied on more degrees of freedom than 
are actually present in the solutions.  This is not necessarily a problem,
but will tend to underestimate our uncertainties.  The results least 
affected will be those based on the lunar Nordtvedt effect --- 
equation~(\ref{Nor_Con1}) --- for which EP-violating terms have been 
included in the solution sets.  Our other results ---
equations~(\ref{Kep_Con1}), (\ref{Kep_Con2}), (\ref{Lag_Con}), 
(\ref{Nor_Con2}) and (\ref{Kep_Con3}) --- may be less robust in comparison.
A fully rigorous future treatment would rely on new statistical fits 
incorporating an independent parameter $\Delta_k$ for each body, rather 
than parametrizing EP violations with a single universal constant.

\section{Application to Kaluza-Klein Gravity} \label{Sec:App}

Let us now see what equations~(\ref{Del_Sun1}) -- (\ref{Del_Ear2}) 
imply for 5D Kaluza-Klein gravity, in which [as we recall from 
equation~(\ref{Del_KK})] $\Delta \approx -b/2$, where $b$ is a 
free parameter which characterizes the departure from flatness of the 
fifth dimension in the vicinity of the massive body.  Our constraints 
on this parameter follow immediately from equations~(\ref{Del_Jup}),
(\ref{Del_Moo}), (\ref{Del_Sun2}) and (\ref{Del_Ear2}), and may be
summarized together as follows
\begin{eqnarray}
|\bs{b}{S}| & \leq & 2 \times 10^{-8} \; ,
   \label{b_Sun} \\
|\bs{b}{J}| & \leq & 2 \times 10^{-6} \; ,
   \label{b_Jup} \\
|\bs{b}{E}| & \leq & 2 \times 10^{-8} \; , 
   \label{b_Ear} \\
|\bs{b}{M}| & \leq & 2 \times 10^{-8} \; .
   \label{b_Moo}
\end{eqnarray}
These upper bounds on $b$ are far more stringent than any obtained so
far by other means.  In particular, the result~(\ref{b_Ear}) for the Earth
is more than three orders of magnitude stronger than that which may be
expected using data on geodetic precession from the upcoming Gravity 
Probe~B satellite \cite{Kal95,Liu00,Buc96}. 
Equation~(\ref{b_Jup}) for Jupiter is more stringent by five orders 
of magnitude than the limits which may be obtained using light deflection
by the giant planet \cite{Liu00,Tre91}. 
And the upper limit~(\ref{b_Sun}) for the Sun is more than {\em six\/} orders
of magnitude tighter than those set so far from the classical tests 
of general relativity (light deflection using long-baseline radio
interferometry, Mercury's perihelion precession, and time delay
using the Mars Viking lander \cite{Kal95,Liu00}).

\section{Conclusions} \label{Sec:Con}

We have looked for the constraints imposed by solar system data
on theories in which the ratio of gravitational to inertial mass differs
from unity by some factor $\Delta$ which may in principle differ from body
to body.  For two objects characterized by $\Delta_1$ and $\Delta_2$,
upper bounds on the sum $|\Delta_1+ c_1 \Delta_2|$ have been
found from Kepler's third law, while the difference $|\Delta_1- c_2 \Delta_2|$
can be constrained by data on the position of Lagrange libration points,
and orbital polarization in the field of a third body (the Nordtvedt effect).
(Here $c_1$ and $c_2$ are known constants.)
Combining these results, we have extracted independent upper limits on 
$\Delta$ for the Sun, Moon, Earth and Jupiter, using experimental data 
on their orbits as well as those of the Trojan asteroids.  In particular,
we find that $\Delta \leq 10^{-8}$ for the Sun, Earth and Moon,
and $10^{-6}$ for Jupiter.

As a test case, we have applied these results to five-dimensional Kaluza-Klein
gravity, in which $\Delta$ can be shown to depend on a parameter $b$ of the
five-dimensional metric which characterizes the curvature of the extra
dimension near the central mass.  Our upper bounds on this parameter are
three to six orders of magnitude stronger than existing limits on 
$b$ from the classical (and geodetic precession) tests of general 
relativity, and confirm earlier conclusions \cite{Ove97,Liu00}
that a fifth dimension, if any, plays no significant role in the 
dynamics of the solar system.

\acknowledgments

The author wishes to acknowledge the invaluable help of an anonymous referee
who corrected several misconceptions and pointed to useful new references.
Thanks also go to Ron Adler, Robb Mann, Wei-Tou Ni, David Santiago, Alex 
Silbergleit and Paul Wesson for discussions; Mihoko Kai for help with the
figures, the National Science and Engineering Research Council of Canada
for financial support, and Francis Everitt and the theory group at 
Gravity Probe~B for hospitality during the time in which part of
this work was carried out.

\newpage

\section*{Figure Captions}

FIG. 1.  Effect of EP violations on the stable three-body configuration
         of Lagrange.

\vspace{1cm}

FIG. 2.  The Nordtvedt effect: two bodies $M_1$ and $M_2$ fall toward
         $M_3$ with different accelerations.

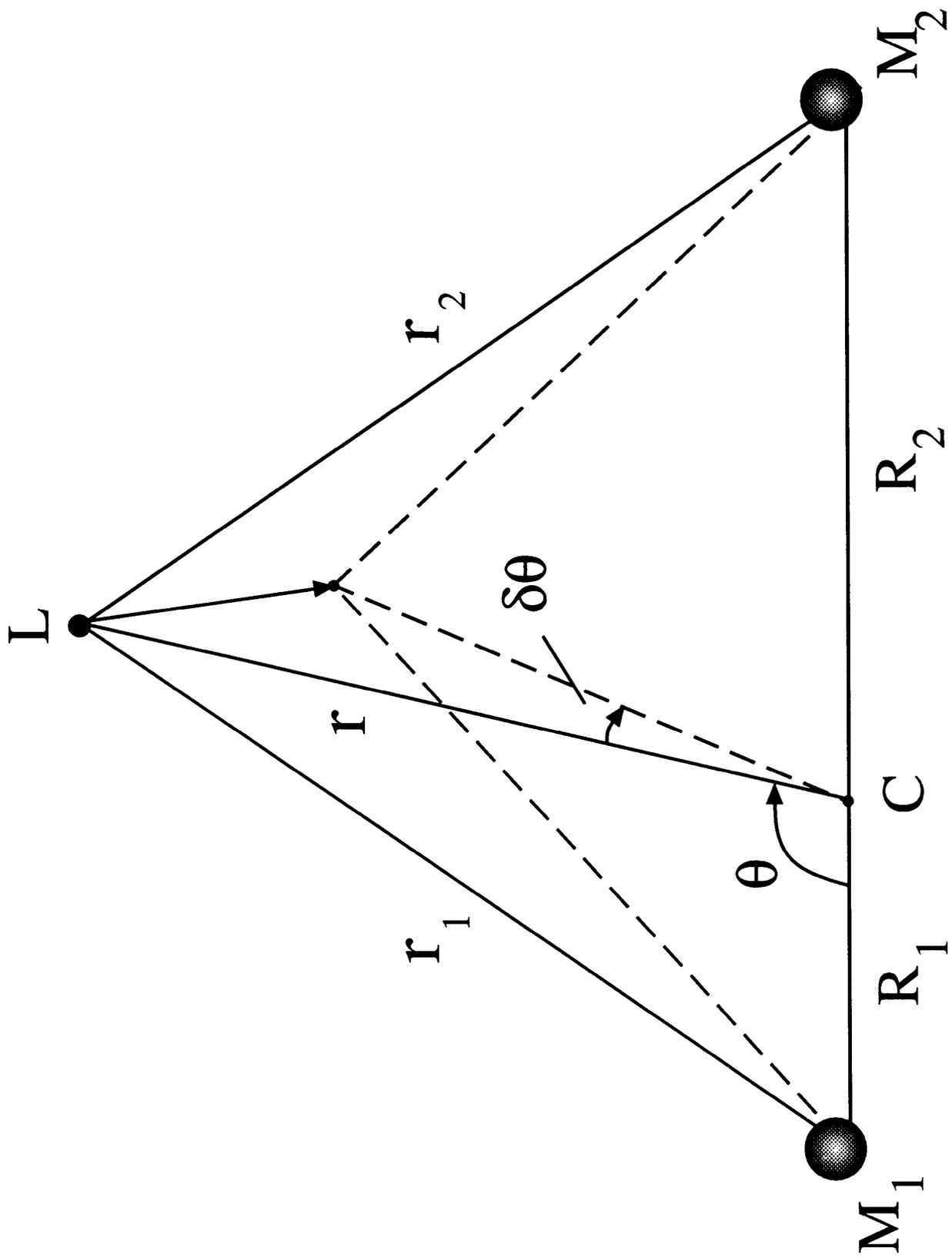
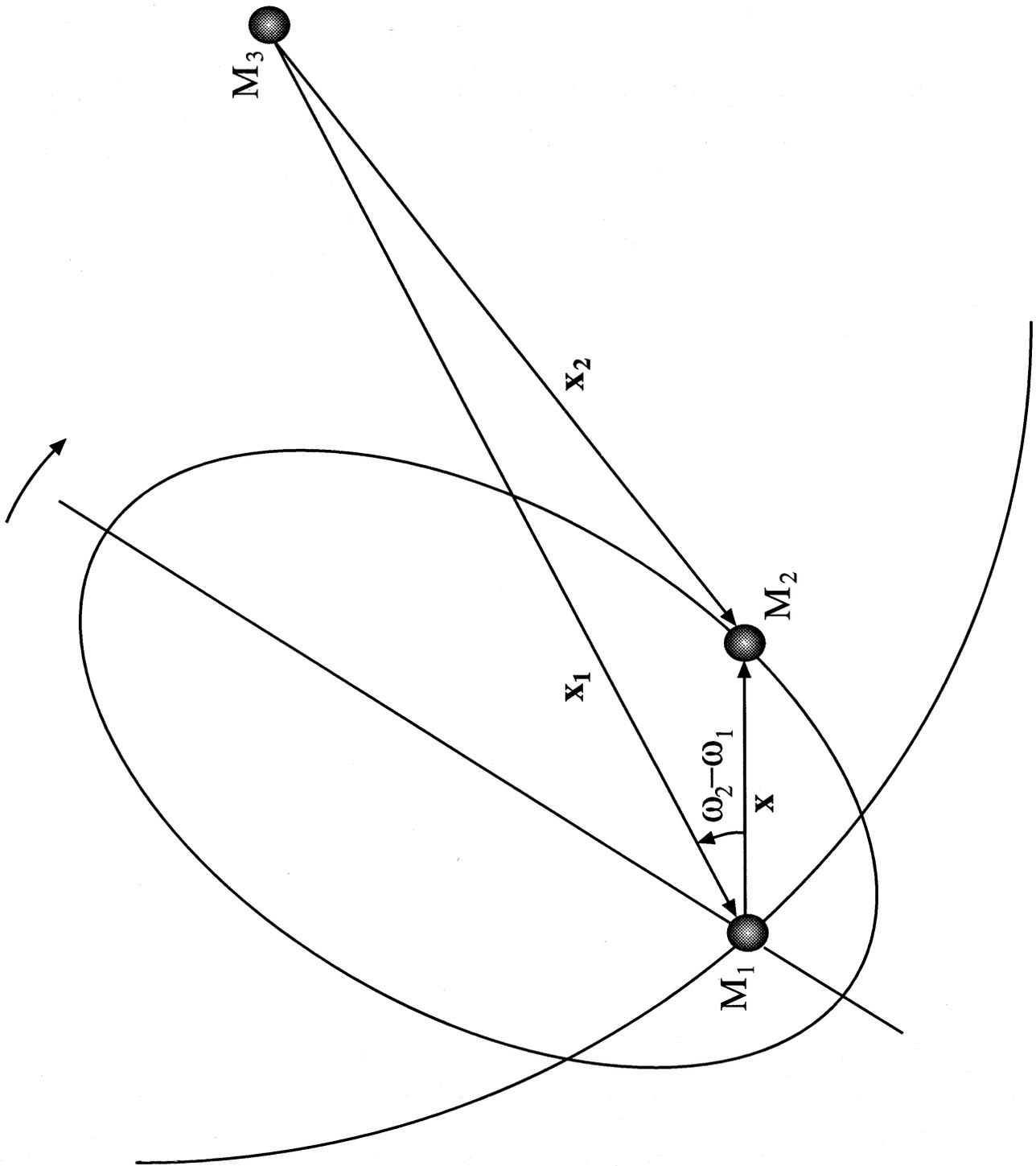
\end{document}